\documentclass[twocolumn,longauth]{aa}  
\usepackage{graphicx}
\usepackage{txfonts}
\usepackage{natbib}
\bibpunct{(}{)}{;}{a}{}{,}
\begin{document}

   \title{An M-dwarf star in the transition disk of Herbig HD 142527}
   \subtitle{Physical parameters and orbital elements}


   \author{S. Lacour\inst{1,2}\and
          B. Biller\inst{3}\and
          A. Cheetham\inst{4}\and
          A. Greenbaum\inst{5}\and
          T. Pearce\inst{6}\and
          S. Marino\inst{6}\and
          P. Tuthill\inst{7}\and
          L. Pueyo\inst{8}\and
          E. E. Mamajek\inst{9}\and
          J. H. Girard\inst{10}\and 
          A. Sivaramakrishnan\inst{8}\and
          M. Bonnefoy\inst{11}\and
          I. Baraffe\inst{9,12}\and
          G. Chauvin\inst{11}\and 
          J. Olofsson\inst{13,14,15}\and
          A. Juhasz\inst{6}\and 
          M.~Benisty\inst{11}\and 
          J.-U. Pott\inst{13}\and 
          A. Sicilia-Aguilar\inst{16}\and 
          T. Henning\inst{13}\and
             A. Cardwell\inst{17}\and
S. Goodsell\inst{18,19}\and
J. R. Graham\inst{20}\and
P.~Hibon\inst{10,17}\and
P. Ingraham\inst{21}\and
Q. Konopacky\inst{22}\and
B. Macintosh\inst{23}\and
R. Oppenheimer\inst{24}\and
M. Perrin\inst{8}\and
F.~Rantakyr\"o\inst{17}\and
N.~Sadakuni\inst{25}\and
S.~Thomas\inst{21}
          }

   \institute{
  LESIA / Observatoire de Paris, PSL, CNRS, UPMC, Universit\'e Paris Diderot, 5 place Jules Janssen, F-92195 Meudon,
France 
 \and
 Cavendish Laboratory, University of Cambridge, JJ Thomson Avenue, Cambridge CB3 0HE, UK
 \and
 Institute for Astronomy, University of Edinburgh, Blackford
Hill View, Edinburgh EH9 3HJ, UK
\and
Observatoire de Gen\`eve, Universit\'e de Gen\`eve, 51 chemin des Maillettes, 1290, Versoix, Switzerland
\and
 Johns Hopkins University Department of Physics and Astronomy, address: 3400 N. Charles St., Baltimore, MD 21218
 \and
 Institute of Astronomy, University of Cambridge, Madingley Road, Cambridge, CB3 0HA, UK
 \and 
 Sydney Institute for Astronomy,
School of Physics,
University of Sydney, NSW 2006, Australia
\and
Space Telescope Science Institute, 3700 San Martin Drive,
Baltimore MD 21218
\and
Astrophysics Group, School of Physics, University of Exeter, Exeter EX4 4QL, UK
 \and 
 European Southern Observatory, Alonso de Cordova 3107, Casilla 19001, Santiago, Chile
 \and
 Univ. Grenoble Alpes, CNRS, IPAG / UMR  5274,  F-38000  Grenoble,
 \and
{\'E}cole Normale Sup{\'e}rieure, Lyon, CRAL (UMR CNRS 5574), Universit{\'e} de Lyon 1, 69007 Lyon, France
\and
 Max Planck Institut f\"ur Astronomie, K\"onigstuhl 17, D-69117 Heidelberg, Germany
 \and
Instituto de F\'isica y Astronom\'ia, 
Universidad de Valpara\'iso, Av. Gran Breta\~na 1111, Playa Ancha, Valpara\'iso, Chile 
 \and
 ICM nucleus on protoplanetary disks, Universidad de Valpara\'iso, Av. Gran Breta\~na 1111, Valpara\'iso, Chile
\and 
School of Physics and Astronomy, University of
St Andrews, North Haugh, KY16 6SS, St Andrews, UK
\and
Gemini Observatory, Casilla 603, La Serena, Chile
\and
 Gemini Observatory, 670 North A'ohoku Place, Hilo, HI 96720, USA
 \and Durham University, Stockton Road, Durham, DH1 3LE, UK
 \and
 Astronomy Department, University of California,  Berkeley CA 94720, USA
 \and
 Large Synoptic Survey Telescope, 950N Cherry Av, Tucson AZ 85719, USA
 \and
 Center for Astrophysics and Space Science, University of California San Diego, La Jolla, CA, 92093, USA
 \and
 Kavli Institute for Particle Astrophysics and Cosmology, Stanford University, Stanford, CA 94305, USA
 \and
 Department of Astrophysics, American Museum of Natural History, New York, NY 10024, USA
 \and NASA / Armstrong Flight Research Center, 2825 East Avenue P, Palmdale, CA 93550, USA
             }

   \date{Received September 15, 1996; accepted March 16, 1997}

 
  \abstract
   {}
   { HD 142527A is one of the most studied Herbig Ae/Be stars
     with a transitional disk, as it has the largest imaged gap in any
     protoplanetary disk: the gas is cleared from 30 to 90 AU.  The HD
     142527 system is also unique in that it has a 
      stellar companion with a
     small mass compared to the mass of the primary star.  This factor
     of $\approx20$ in mass ratio between the two objects makes this
     binary system different from any other YSO. The HD142527 system could therefore provide a valuable test bed for understanding the
impact of a lower mass companion on disk structure.
This low-mass stellar object
     may be responsible for both the gap and dust trapping
     observed by ALMA at longer distances.
     }
   {
   We observed this system with the NACO and GPI instruments using the aperture masking technique.
   Aperture masking is ideal for providing high dynamic
   range even at very small angular separations. 
   We present the spectral energy distribution (SED) for HD 142527A and B. Brightness of the companion is now known from
   the $R$ band up to the $M'$ band. We also followed the orbital motion of HD
   142527B over a period of more than two years. }
 { The SED of the companion is compatible with a $T=3000\pm100\,$K object in addition
   to a 1700\,K blackbody environment (likely a circum-secondary disk). From evolution models, we
   find that it is compatible with an object of mass
   $0.13\pm0.03\,M_{\sun}$, radius $0.90\pm0.15\,R_{\sun}$, and age
   $1.0^{+1.0}_{-0.75}\,$Myr. This age is significantly younger than the age previously estimated for HD 142527A.
   Computations to constrain the orbital parameters found a semimajor axis of
   $140^{+120}_{-70}$\,mas, an eccentricity of $0.5 \pm 0.2$, an
     inclination of $125 \pm 15$ degrees, and a position angle of the
     right ascending node of $-5 \pm 40$ degrees. 
     Inclination and position angle of the ascending node are in agreement with an orbit coplanar with the inner disk, not coplanar with the outer disk.
     Despite its high eccentricity, it is unlikely that HD 142527B is responsible for truncating the inner edge of the outer disk. }
   {} 


   \maketitle
%

\section{Introduction}

During the process of planet formation and disk dissipation, primordial disks begin in an optically thick state with significant emission at infrared wavelengths. As the disks clear out, they pass through an intermediate transitional stage marked by a drop in near- or mid-infrared (IR) emission that indicates the presence of an annular disk gap.
These gaps can form through several mechanisms, such as photoevaporation, truncation due to a binary companion, or the presence of a forming planet. This last pathway has led to transitional disks being the subject of close study, as they may provide valuable insights into the planet formation process.

HD 142527A, a young stellar object (YSO), is one of the most  studied Herbig Ae/Be stars with a
transitional disk. With dust and some gases cleared from 30 to 90 AU,
  it has the largest imaged gap in any
protoplanetary disk. 
According to \citet{2014ApJ...790...21M}, the age, distance, and
mass for this star are $5 \pm 1.5$\,Myr, $140 \pm 20$\,pc and $2.0 \pm
0.3\,M_{\sun}$, respectively.   \citet{2012ApJ...753L..38B} discovered a low-mass stellar companion 
($\sim$0.2 M$_{\sun}$) at $\approx 12\,$AU
from the star within the gap.   This companion was  
confirmed in the $R$ band by \citet{2014ApJ...781L..30C}. Eccentric orbit or not, it may have played a vital role in carving the large
gap in this system.

HD 142527A is also notable for its diversity of disk structure.
Recent imaging with the Atacama Large Millimeter Array (ALMA) of the outer disk
\citep{2013Natur.493..191C,2015arXiv150507743C} reveals
large asymmetrical structures composed of millimeter-size grains 
\citep[similar to the rare horseshoe structure also seen around the
young Herbig star WLY 2-48 by][]{2013Sci...340.1199V}. 
These structures are thought to be signposts of
density variations possibly caused by planetary formation \citep{2011ARA&A..49...67W}.  
Scattering of the stellar light in the near-infrared also reveals
structures at the surface of the outer disk  \citep{2014ApJ...781...87A}.

The \object{HD 142527} system provides a valuable test bed for understanding the
impact of a lower mass companion on disk structure.  As opposed to
other young stellar binaries, which generally have nearly equal masses, the HD
142527 system has a more extreme mass ratio (a factor of $\sim$20).  
Thus, we can study in situ the effects of
this companion, which may be responsible for both the gap and dust
trapping observed further out in this disk.  
However, initial constraints on the mass and other properties of the companion are still vague and at its present apparent separation it does not appear to be responsible for clearing the large disk gap. Many studies have invoked the possibility of forming giant planets on wide orbits exterior to the companion, making HD 142527 an exciting target for direct imaging observations. Constraining the mass and orbital parameters of the low-mass companion is vital to determining the origin of the observed disk gap.

We have been conducting ongoing
 orbital monitoring to constrain the companion properties.
This paper reports multiple aperture masking observations of
HD142527B. The interferometric nature of the technique permits
precise measurements of the contrast ratio and separation, even 
though the companion lies at the diffraction limit of the telescope. 
In Section~\ref{secObs}, we present the new sparse aperture masking (SAM) observations. 
In Section~\ref{secA}, we review the basic parameters of HD142527A.
In Section~\ref{secSED}, we report and discuss the SED of the
companion from visible to mid-infrared (5$\,\mu$m) light. 
In Section~\ref{secOrb}, we present the orbital parameters of 
the companion. In Section~\ref{secCon}, we discuss the results.

   \begin{figure}
   \centering
  \includegraphics[height=4.9cm]{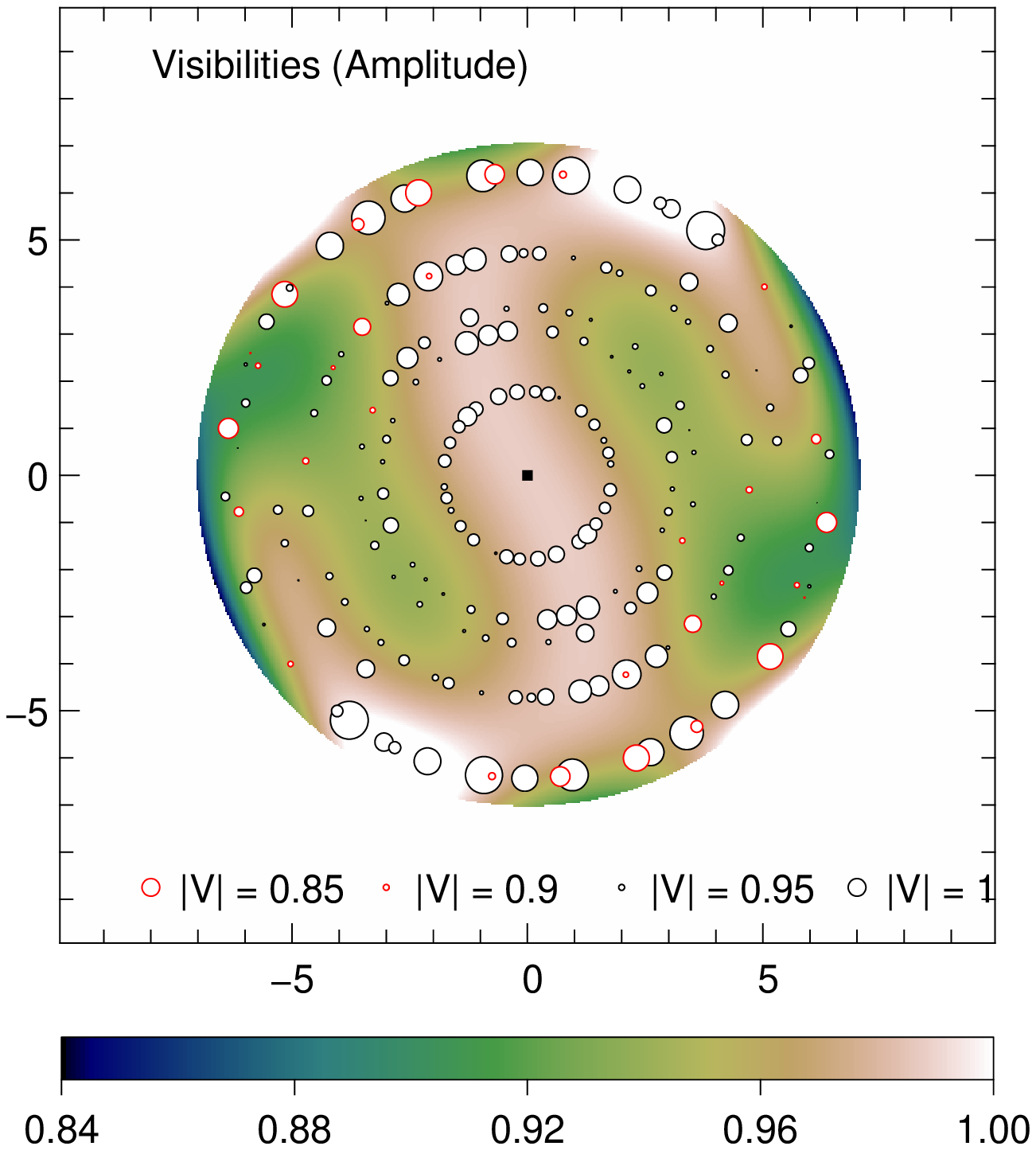}
  \includegraphics[height=4.9cm]{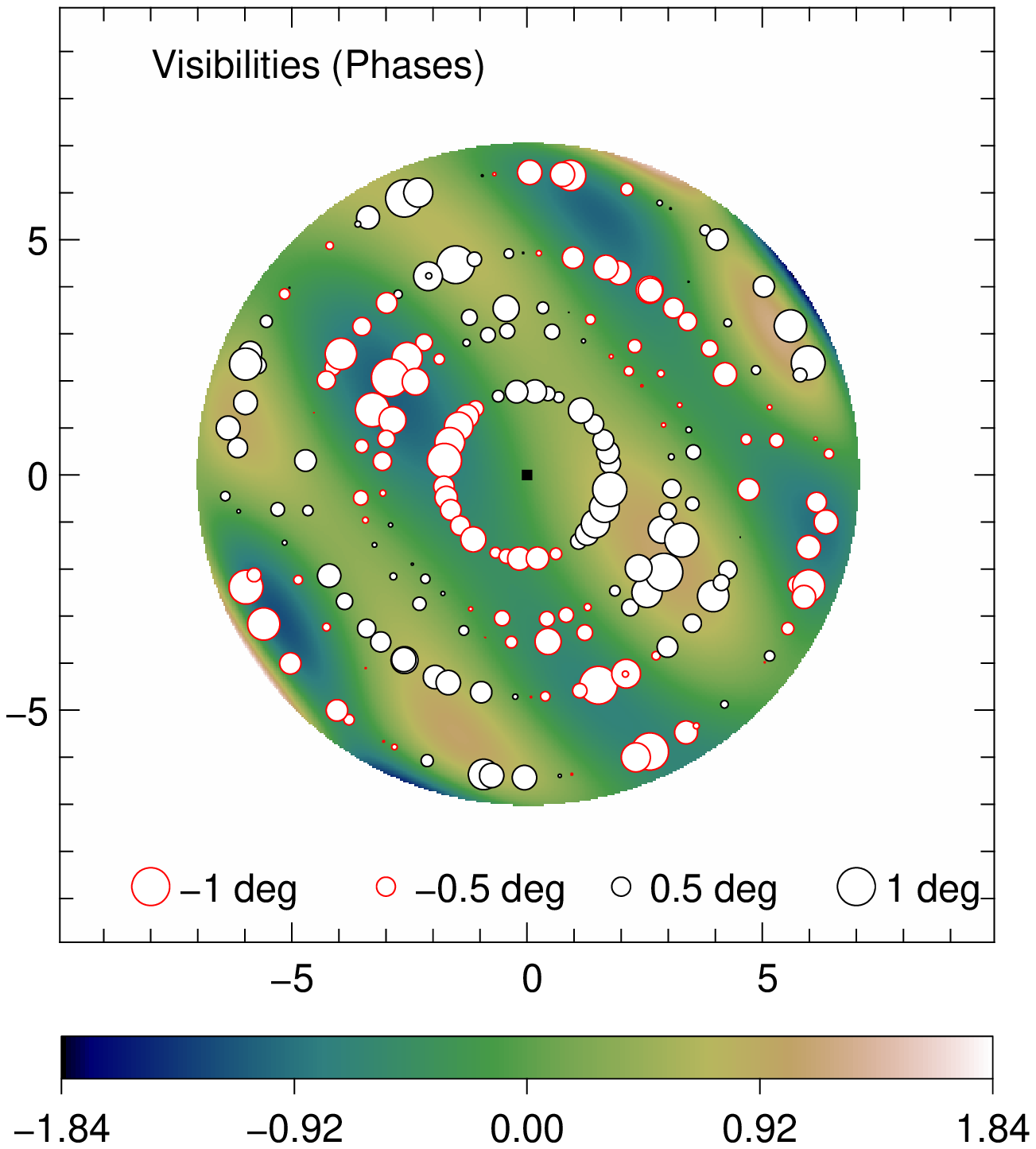}
      \caption{SAM K band dataset. The phase and amplitude in the Fourier domain are represented by the size and color of the open circles. The units are normalized visibility for the amplitudes (left panel) and degrees for the phases (right panel). The color continuum below the data point are obtained by fitting Zernike polynomials to the data. The diagonal stripes observed on the phases are typical of a binary system.}
         \label{figData}
   \end{figure}
 
\section{Observations and data reduction}
\label{secObs}

\subsection{Principle}

   \begin{table}
      \caption[]{Observations and contrast ratio}
         \label{tbObs}
         \centering
\begin{tabular}{l c c c}
\hline\hline
Instrument & Spectral Band & Date & $\Delta$mag \\
\hline 
MagAO$^{(1)}$ &$H\alpha$ cont & 11 April 2013&  $7.5\pm0.25$ \\
GPI$^{(2)}$(Imaging) & Y Band & 25 April 2014&  $3.2\pm0.5$ \\
\hline 
 NACO (SAM)    & $J$ Band &17 March 2013      &$5.0\pm0.6$ \\
   &    &14 July 2013     &$5.0\pm0.2$ \\
  GPI (SAM) & & 12 May 2014  &$4.63\pm0.04$ \\
\hline 
 NACO (SAM)     & $H$ Band &11 March 2012 &$4.5\pm0.5$\\
\hline 
  NACO (SAM)      & $Ks$ Band   & 11 March 2012     &$4.80\pm0.09$\\
   & &17 March 2013      &$4.78\pm0.07$ \\
 & &  14 July 2013       &$4.67\pm0.05$ \\
\hline 
  NACO (SAM)     & $L'$   Band   &   11 March 2012    &$5.16\pm0.11$\\
    &    &17 March 2013    &$5.36\pm0.09$ \\
\hline 
 NACO (SAM)  & $M'$ Band   &17 March 2013       &$5.77\pm0.34$ \\
   & & 14 July 2013     &$5.80\pm0.16$ \\
\hline
\end{tabular}
\tablebib{
(1) \citet{2014ApJ...781L..30C}; (2) \citet{2014ApJ...791L..37R} 
}
\end{table}

Nonredundant aperture masking (NRM), also called sparse aperture
masking (SAM) allows a single aperture telescope to be used as a
Fizeau interferometer. This is accomplished by dividing the
aperture of the telescope into multiple subapertures using 
a mask in the pupil plane of the telescope, then allowing light
from the subapertures to interfere.
The point spread 
function (PSF) is thus transformed into a pattern made of multiple
fringes.  Measuring the amplitude and phase of
these fringes allows us to retrieve information to the
diffraction limit of the telescope and beyond.  The use of closure 
phases with SAM also enables high contrast capability, 
 as closure phases are an
estimator robust to atmospheric and optical aberrations \citep{1976JOSA...66.1236K}.

\subsection{Datasets and image reconstruction}

   \begin{figure}
   \centering
  \includegraphics[height=6.5cm]{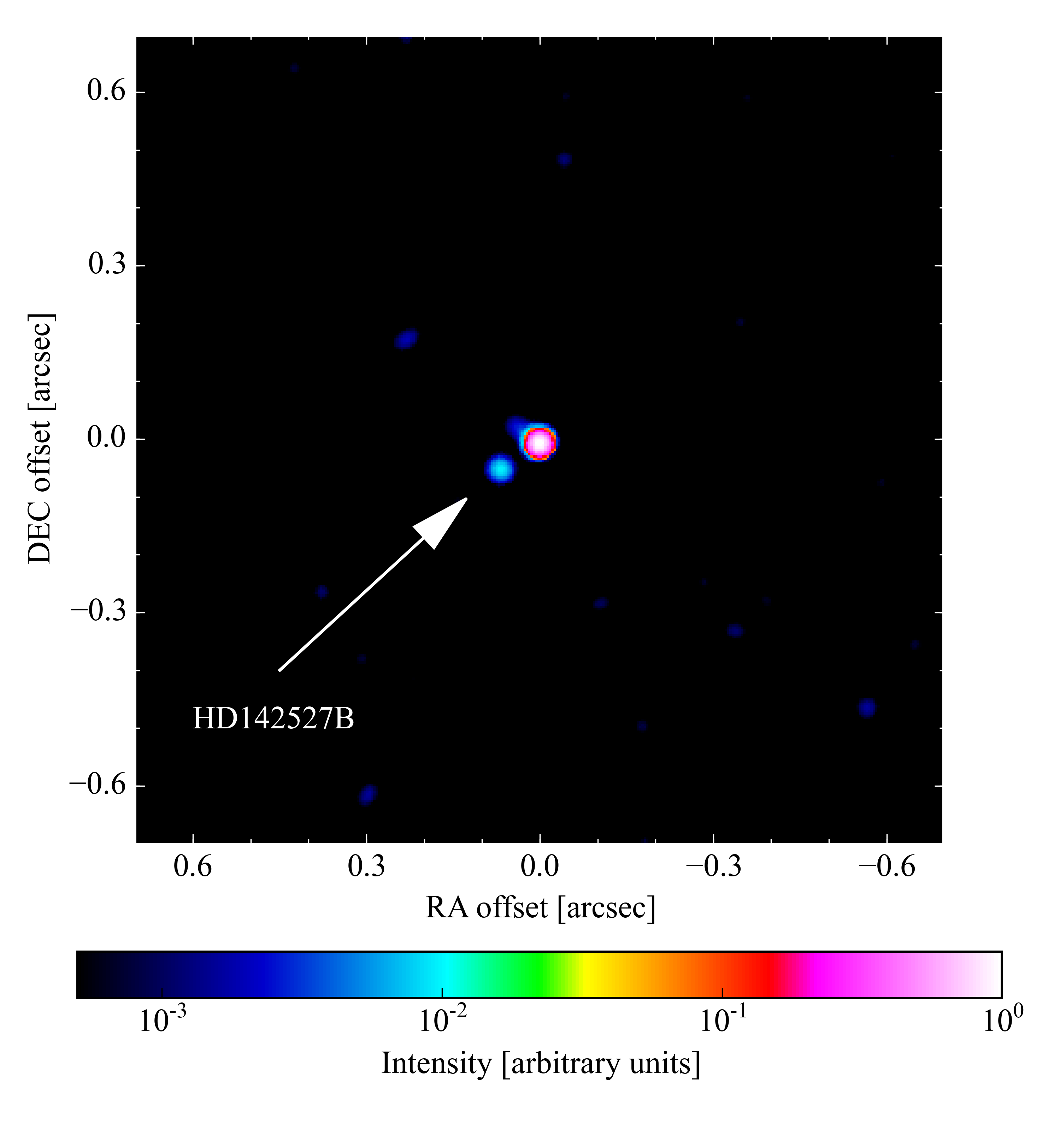}
      \caption{SAM K band image reconstruction generated using the MIRA reconstruction software \citep{2008SPIE.7013E..1IT}.}
         \label{figImages}
   \end{figure}

We observed HD142527 with two instruments: NACO at the VLT, and the Gemini Planet Imager (GPI)
on the Gemini South telescope. Both instruments are equipped with NRM masks. We used a 7-hole on NACO \citep{2010SPIE.7735E..1OT} and the 10-hole on GPI \citep{2014SPIE.9147E..7BG}. 

New NACO data were
obtained in March 2013 and July 2013. All NACO data were
reduced with the SAMP pipeline \citep{2011A&A...532A..72L}.
The two calibrators HD\,142695 and HD\,144350 were interleaved with the observations.
We  reanalyzed the March 2012 NACO data from \citet{2012ApJ...753L..38B}  to remove HD\,142384 as a calibrator, which turned out to be a close binary system \citep{2014arXiv1408.3227L}. The new reduction also benefits from
recent updates to the SAMP pipeline to include a
post-processing atmospheric dispersion corrector. 

The GPI data were obtained in May 2014 in the $J$ band. Calibrator HD142695 was observed immediately afterward to remove instrumental effects.
The GPI data reduction pipeline \citep[][and references therein]{2014SPIE.9147E..3JP} was used to extract images in 37 wavelength channels, of which 17 are expected to be independent. Measured uncertainties were scaled to take this effect into account. 
The GPI pipeline corrects for flexure, bad pixels, and distortion in the two-dimensional (2D) frame. Argon lamp arcs taken right before the observations were used to calibrate wavelength and correct for shifts in position of microspectra due to flexure
\citep{2014SPIE.9147E..7HW}. The data were dark subtracted and assembled into data cubes. 
The data cubes were then processed using the aperture masking pipeline developed at Sydney University with the closure phases and visibilities measured in different wavelength channels treated independently. 
These data were fit with a three-parameter binary model (separation, position angle, and contrast ratio), with the contrast ratio assumed to be constant across the wavelength channels.

To illustrate the typical data produced by NRM, 
the Ks band data obtained in July 2013 are presented
in Fig~\ref{figData}. The $u$-$v$ frequency domain is limited by the
diameter of the 8-meter telescope. The amplitudes (visibilities) are
normalized by the contrast observed on the calibrators. The right
panel shows the phase derived from the closure phase. The closure
phase to phase relation was established by singular value
decomposition of the closure phase-to-phase matrix. Missing phase
parameters (piston, tip, tilt, etc...) are set to zero.

We used closure phases and amplitudes to reconstruct a high
resolution deconvolved image with the MIRA software
\citep{2008SPIE.7013E..1IT}. The resulting high resolution 
deconvolved image is shown for the July 2013 Ks band dataset
in Fig.~\ref{figImages}.  We performed the image synthesis using a
nonparametric least-squares modeling technique over the amplitudes
and closure phases, and adding a regularization term called ``{\it
  xsmooth}'' ($\ell_2 - \ell_1$ smoothness constraint).  A
point-source model image was used as seed for the image
reconstruction. We found that the companion detection is robust to
different regularization methods.

\subsection{Companion fitting}

Parameters describing the companion were extracted through model fitting
to the closure phase data only, discarding the visibilities. At each epoch
of observation, a single position was simultaneously fitted to all
spectral bands to better constrain the flux ratio at
long wavelengths.  The model fitted to the closure phase is the following:
\begin{eqnarray}
CP({\vec u}, {\vec v}, \lambda) & = &    \arg \left[1+  \rho_\lambda \exp(2 i \pi   ( \vec  \alpha \cdot \vec u)/\lambda)\right] \nonumber\\
&& + \arg \left[1 + \rho_\lambda \exp(2 i \pi ( \vec \alpha \cdot \vec {v})/\lambda)\right]\nonumber \\ && + \arg \left[1 + \rho_\lambda \exp(-2 i \pi \vec \alpha \cdot (\vec{u}+\vec{v})/\lambda)\right]
,\end{eqnarray}
where $\lambda$ is the wavelength, and $(\vec u , \vec {v})$ the two baseline-vectors that compose the closure phase triangle. The free parameters of the fit are: i) the separation vector $\vec \alpha$ and ii) $\rho_\lambda$, the flux of the secondary with respect to the primary. The separation $\vec \alpha$ is wavelength independent, but the flux of the secondary depends on the filter bandpass (from $J$ band to $M'$ band). This procedure helps mitigate the problem of limited angular resolution at longer 
wavelengths. 

The best solution for $\vec \alpha$ is achieved by taking the minimum value of the 2D $\chi^2$ map. The map is obtained by regression over the free parameters $\rho_\lambda$. The map is then normalized to a reduced $\chi^2$ of one, and the three sigma errors are determined from the area below which the reduced $\chi^2$ is less than 10. An example of such a map is given in the central panel of Fig~1 in \citet{2012ApJ...753L..38B}.

Finally, we determined the
flux ratio  for each filter,  $\rho_\lambda$, adopting the
position (including its error) obtained at each epoch. The flux ratios
between the central star (including the unresolved part of the inner
disk) and the companion are presented in Table~\ref{tbObs}. Unless stated otherwise, errors mentioned in this paper are 1 sigma.

\section{HD142527A}
\label{secA}

\subsection{Distance, age, and mass}
\label{secdist}

We adopt a distance of 140\,$\pm$\,20 pc based on the arguments made
by \citet{2014ApJ...790...21M}, namely that the star is near the Lupus
IV cloud, and shares the proper motion and radial velocity of
neighboring Sco-Cen stars \citep[e.g.,][]{2008hsf2.book..235P}.  The proper motion of the star \citep[$\mu_{\alpha}$ = -17.2\,mas/yr, $\mu_{\delta}$ =
  -18.0\,mas/yr ;][]{2007A&A...474..653V} is similar to that of overall
proper motion of stars associated with the Lupus clouds
\citep[$\mu_{\alpha}$ = -16.0\,mas/yr, $\mu_{\delta}$ = -21.7\,mas/yr
  ;][]{2013A&A...558A..77G}.  The star is likely to be co-distant with either the
Lupus clouds \citep[][]{2013A&A...558A..77G} or Upper Cen-Lup subgroup of Sco-Cen
\citep[$\sim$142 pc][]{1999AJ....117..354D}, which surrounds the Lupus clouds. 

\citet{2014ApJ...790...21M} used an X-Shooter spectrum to constrain the temperature and the \citet{2001ApJS..136..417Y} tracks to estimate the age and mass of HD142527A. These do provide a  dense grid of tracks for masses of $\sim2 M_{\sun}$. However evolution models can be prone to systematic error in age and mass, especially for young stars. Therefore, we run the star parameters through different isochrone pre-main-sequence track models.
When we adopt the $T_{\rm eff}=6550\pm 100\,K$ and luminosity $L=16.3\pm4.5\,L_{\sun}$ from \citet{2014ApJ...790...21M}, we get the following age \& mass pairs:

\citet{1997MmSAI..68..807D} tracks:
\begin{itemize}
\item log(age/yr) = 6.69+0.07-0.15 ($1\sigma$) +0.13-0.39 ($2\sigma$)
\item age(Myr)    = 4.9+0.8-1.4 ($1\sigma$) +1.8-2.9 ($2\sigma$)
\item Mass($M_{\sun}$ )  = 2.08+0.24-0.13 ($1\sigma$) +0.72-0.25 ($2\sigma$)
\end{itemize}

\citet{2000A&A...358..593S} tracks:
\begin{itemize}
\item log(age/yr) = 6.80+0.11-0.12 ($1\sigma$) +0.21-0.25($2\sigma$)
\item age(Myr)    = 6.3+1.7-1.5 ($1\sigma$) +3.9-2.7($2\sigma$)
\item Mass($M_{\sun}$ )  = 2.01+0.21-0.18 ($1\sigma$) +0.48-0.33 ($2\sigma$)
\end{itemize}

Yonsei-Yale \citep{2003ApJS..144..259Y} tracks:
\begin{itemize}
\item log(age/yr) = 6.70+0.11-0.13 ($1\sigma$) +0.24-0.27 ($2\sigma$)
\item age(Myr)    = 5.0+1.5-1.3 ($1\sigma$) +3.7-2.3 ($2\sigma$)
\item Mass($M_{\sun}$ )  = 2.07+0.21-0.18 ($2\sigma$) +0.46-0.37 ($2\sigma$)
.\end{itemize}

We adopt a mass of $2.05\,M_{\sun}$, with a statistical and observational
error of $\pm0.2\,M_{\sun}$. We add a systematic error of $\pm0.04\,M_{\sun}$ from the tracks, assuming these three sets of tracks are sampling our ignorance in composition in physics. However, we conclude that the values given by \citet{2014ApJ...790...21M} are a good approximation: a mass of $2.0\pm0.3\,M_{\sun}$ and an age of $5.0\pm 1.5\,$Myr. 

   \begin{table}
      \caption[]{Visible and near-infrared fluxes of HD142527B}
         \label{tbFlux}
         \centering
\begin{tabular}{l c c c c}
\hline\hline
Band  & $\Delta$mag & HD142527A   & HD142527B  &  HD142527B\\
  &  &  [Jy] &  [mJy]  & [Abs. Mag.]\\
\hline 
$R_{cont}$ &  $7.5 \pm 0.25$\tablefootmark{(1)}         &  $1.8\pm 0.2$\tablefootmark{(3)}   & $1.8 \pm 0.3$  & $9.3\pm 0.3$\\
$J$ &      $4.63 \pm 0.04$                      & $3.8\pm0.1$\tablefootmark{(2)}& $53 \pm 2$     & $5.13\pm 0.05$   \\
$H$ &      $4.5 \pm 0.5$                & $4.6\pm0.1$\tablefootmark{(2)}&  $74 \pm 32$ & $4.6 \pm 0.5$ \\
$K_S$ &      $4.72 \pm 0.04$            & $5.6\pm0.1$\tablefootmark{(2)}& $72 \pm 3$ & $4.1 \pm 0.1$\\
$L' $&      $ 5.26 \pm 0.08$            & $7.4\pm0.3$\tablefootmark{(2)} & $58 \pm 5$  & $3.4 \pm 0.1$ \\
$M'$ &     $5.79 \pm 0.16$              & $6.7\pm0.3$\tablefootmark{(2)}& $32 \pm 5$  & $3.5 \pm 0.2$ \\
\hline
\end{tabular}
\tablebib{
(1) \citet{2014ApJ...781L..30C}
(2) \citet{2011A&A...528A..91V}
(3) from SED using \citet{2004astro.ph..5087C} stellar models
}
\tablefoot{
Absolute magnitudes are obtained assuming a distance of 140pc and dereddened from an $A_V=0.6$ dust absorption.
}
\end{table}

\subsection{Spectral energy distribution}

   \begin{figure}
   \centering
  \includegraphics[height=7cm]{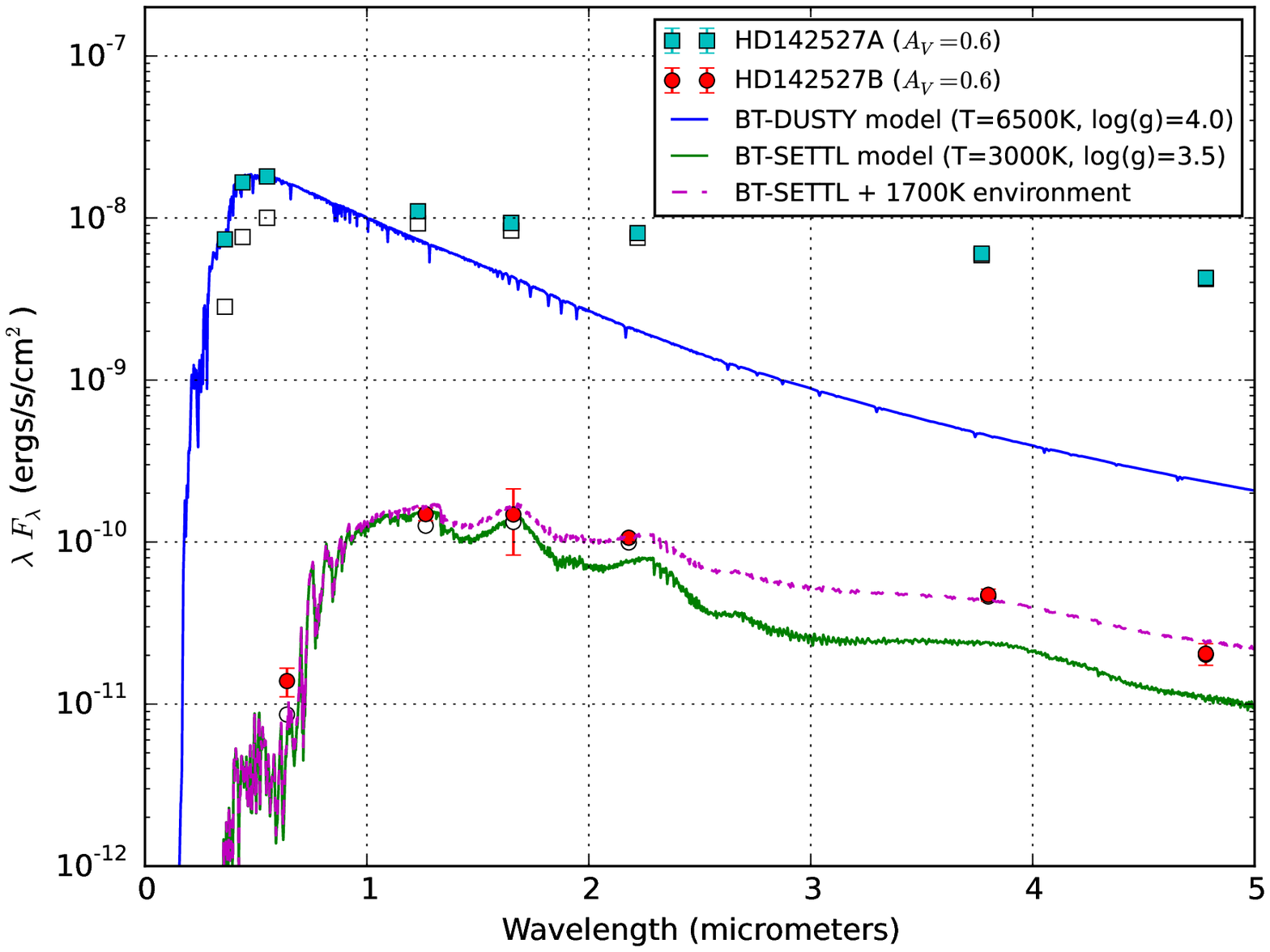}
  \includegraphics[height=7cm]{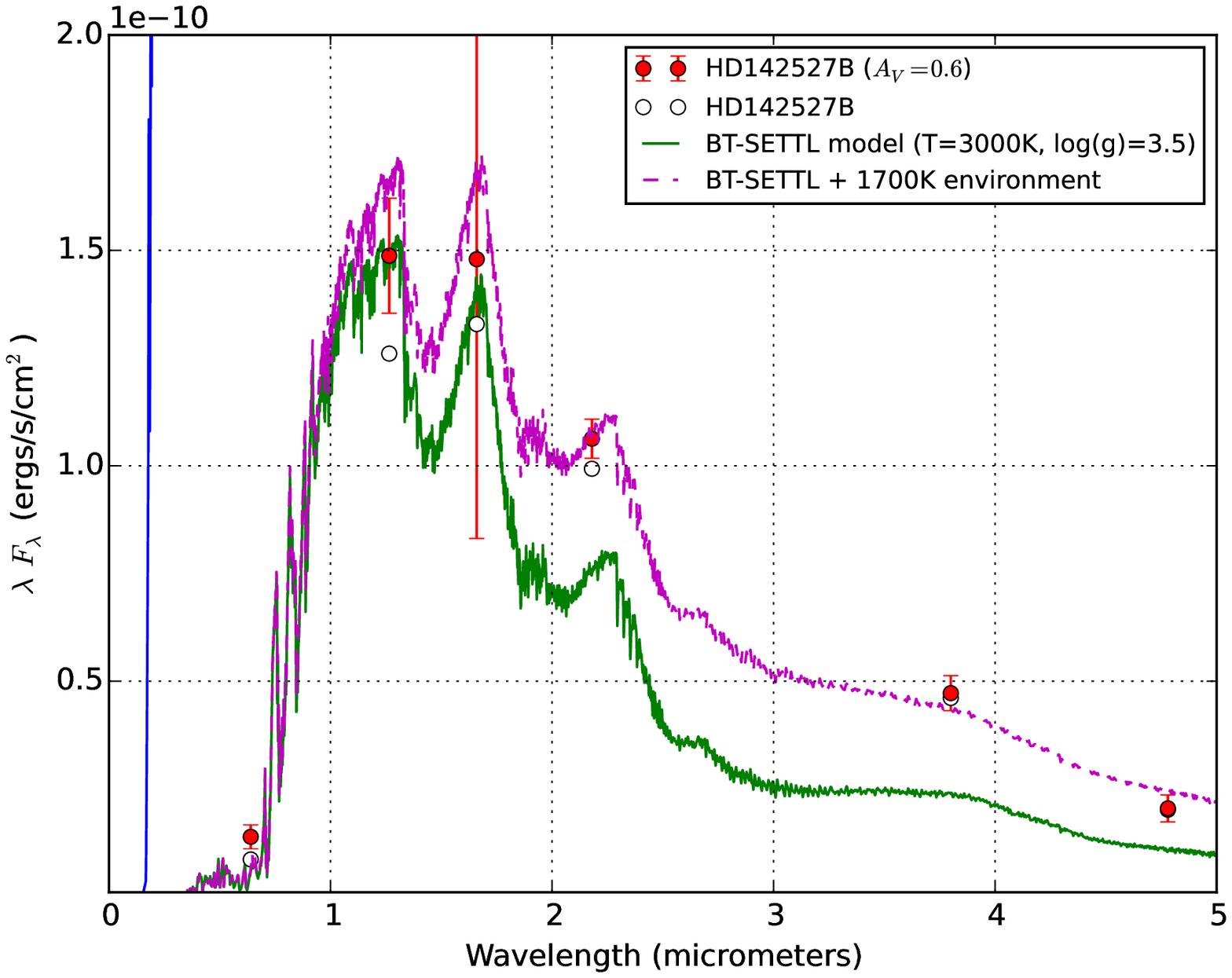}
      \caption{Spectral energy distribution of HD142527 A \& B. The empty symbols are uncorrected for reddening. The filled symbols include
 correction for a dust opacity of visible absorption magnitude $A_V=0.6$. Blue and green curves are stellar models of both stars from \citet{2012RSPTA.370.2765A}.}
         \label{figSED}
   \end{figure}

We plot the SED for HD 142527A in Fig.~\ref{figSED}.  We computed the SED from the fluxes reported by
\citet{2011A&A...528A..91V}, as compiled in Table~\ref{tbFlux}.   
Dust reddening was obtained from dust models by
\citet{2001ApJ...548..296W} ($R_V$=3.1). Following
\citet{2011A&A...528A..91V}, we used a visible extinction
$A_V=0.6$\,mag, which includes the extinction by the interstellar medium and by the circumstellar material. 
This is in rough agreement with the excess color estimated by \citet{1998A&A...331..211M} and \citet{2014ApJ...790...21M} of $E(B-V)= 0.33$ and 0.25\,mag, respectively .

We obtain the blue spectral curve presented in Fig.~\ref{figSED} by
adjusting a synthetic model of stellar atmosphere to the HD142527A
dereddened fluxes.  We used a \citet{2012RSPTA.370.2765A} DUSTY model
of temperature 6500\,K and density log(g)$=4.0$. 
The main objective was to determine the $R$ band continuum emission of 
HD142527A. The result is within the error bar of using a temperature model of 6550\,K and log(g) of 3.75 as derived by \citet{2014ApJ...790...21M}. The $R$ band continuum flux is reported in Table~\ref{tbFlux}.

\section{Physical parameters of HD142527B}
\label{secSED}

   \begin{figure}
   \centering
  \includegraphics[width=9.2cm]{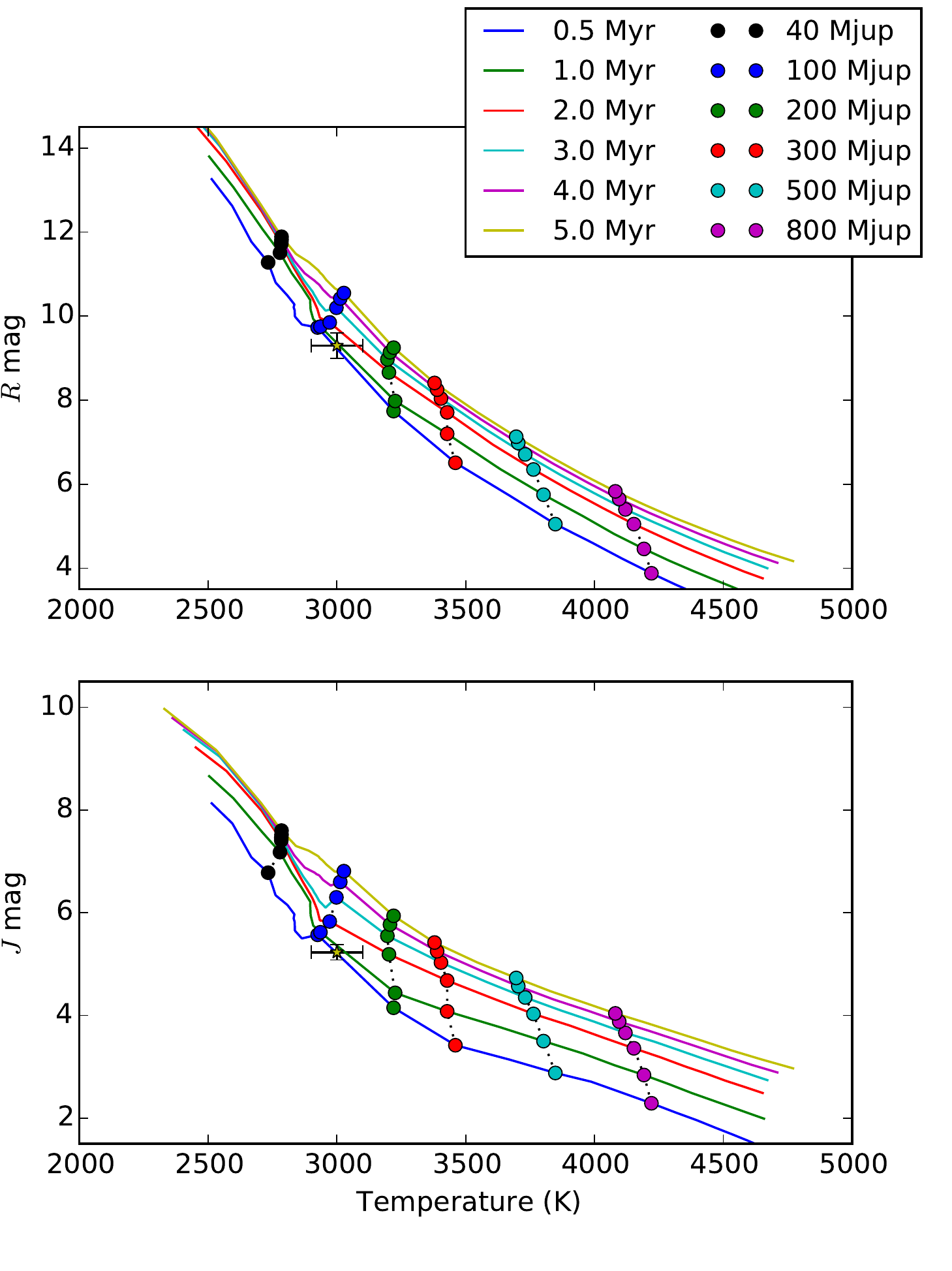}
      \caption{Isocontours of masses and ages as a function of temperature and absolute magnitude, according to evolutionary models from \citet{2015A&A...577A..42B}. The central asterisk correspond to HD142527B. HD14252A, with an effective temperature on the order of 6500\,K, lies outside the plot and outside the temperature range of the model.}
         \label{figEv}
   \end{figure}

\subsection{Spectral energy distribution}

We fit the SED of HD142527B with a combination of stellar emission and dust emission from the circumsecondary disk. The dust emission is approximated by a blackbody of
temperature 1700\,K, which is close to the sublimation limit of dust
particles.  We used the BT-SETTL models computed by
\citet{2012RSPTA.370.2765A}\footnote{http://perso.ens-lyon.fr/france.allard/} to fit the stellar component of the SED. 
The parameters of this model are a
temperature of 3000\,K and a surface gravity of $log(g)=3.5$. 

We find that the dust component modeled as a 
1700\,K blackbody emission primarily fits the longer wavelengths
in the SED, while the J and R band emission is emitted almost 
exclusively ($>90\%$) from the stellar surface. Thus,
we can derive the temperature of the stellar surface
from the J-R color of the object. 

The absolute magnitudes are reported in the right column of
Table~\ref{tbFlux}. These values are computed using a distance
of 140\,pc and dereddened using the algorithm determined by
\citep{1989ApJ...345..245C}. In the following study, we adopt a $J$ band absolute magnitude of $5.23\pm0.1\,$mag. The absolute magnitude includes an additional 0.1\,mag, corresponding to $\approx10\%$, to remove circumstellar emission. Dust
emission is assumed to be negligible in the $R$ band.

Hence, we obtain $(R-J)_{\rm
  HD142527B}=4.17\pm 0.3$\,mag, where the error is dominated by the $R$ band
contrast ratio uncertainties. According to the catalog of stellar
surface colors from \citet{2012RSPTA.370.2765A}, we derive an
effective temperature of $T_{\rm HD142527B}=3000\pm 100$\,K, assuming log$(g)=3.5$.

\subsection{Estimation of mass, age, and radius}

The ratio between the radius of HD142527A and HD142527B can be
estimated from the relative magnitude and effective temperatures
of the stars. It is $R_{\rm HD142527A}/R_{\rm HD142527B}
=2.8\pm0.2$ if we assume an effective temperature of 6500\,K for the primary and 3000\,K for the secondary.

The mass, age, and absolute radius can be obtained from
the absolute magnitudes, however, it requires prior knowledge of the evolution process of the star.
Thus, we used the standard evolutionary model from
\citet{2015A&A...577A..42B}. This model does not assume any accretion
during the evolution of the system. We plotted in
Fig.~\ref{figEv} the isochrome and isomass of the star as a function of the temperature and $R$ and $J$ absolute magnitudes. According to these models, a suitable set of parameters are $M_{\rm
  HD142527B}=0.13\pm0.03\,M_{\sun}$, $R_{\rm
  HD142527B}=0.90\pm0.15\,R_{\sun}$ and an age of
$1.0^{+1.0}_{-0.75}\,$Myr. All errors are 1 sigma.

\section{Orbital elements of HD142527B}
\label{secOrb}

\subsection{Small arc analysis}

   \begin{figure}
   \centering
  \includegraphics[height=11.5cm]{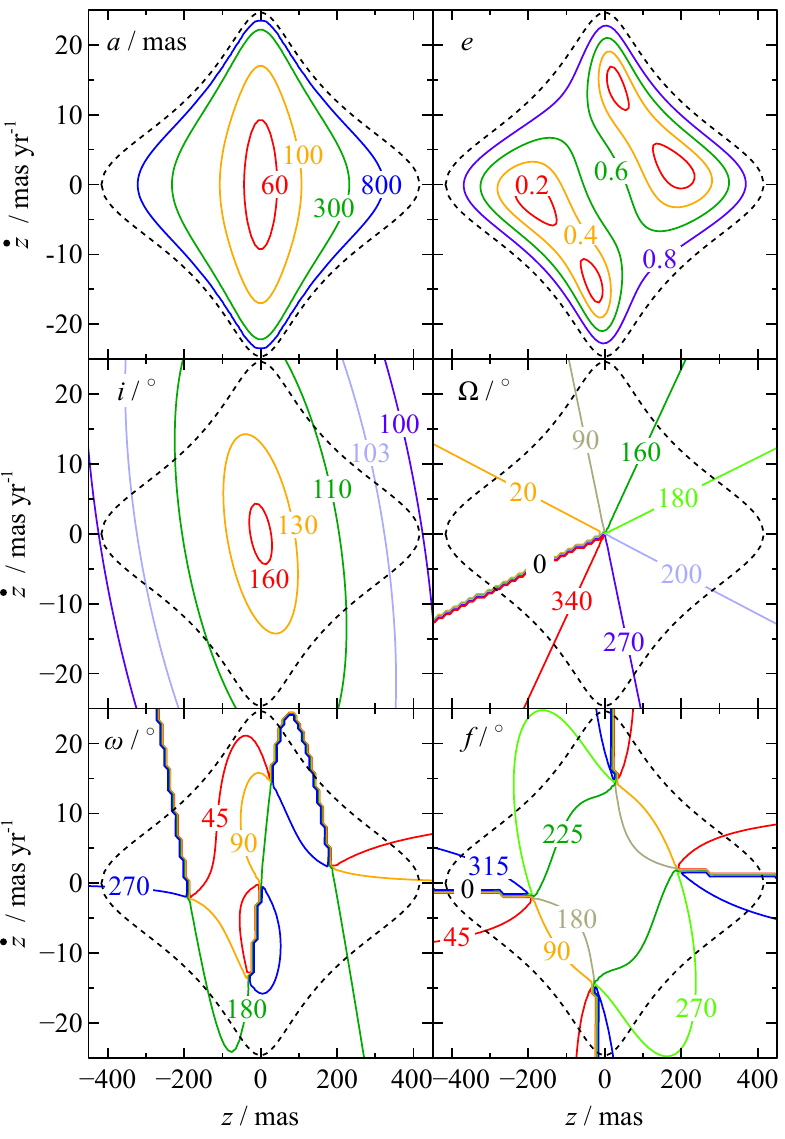}
      \caption{Orbital elements of the system for different $z$ and $\dot z$ values. The contours are obtained by assuming a total system mass of $M_{\rm AB}=2.2\,M_{\sun}$ and a distance to the system of $d=140$\,pc.}
         \label{z_zdot}
   \end{figure}

   \begin{figure}
   \centering
  \includegraphics[height=7.1cm]{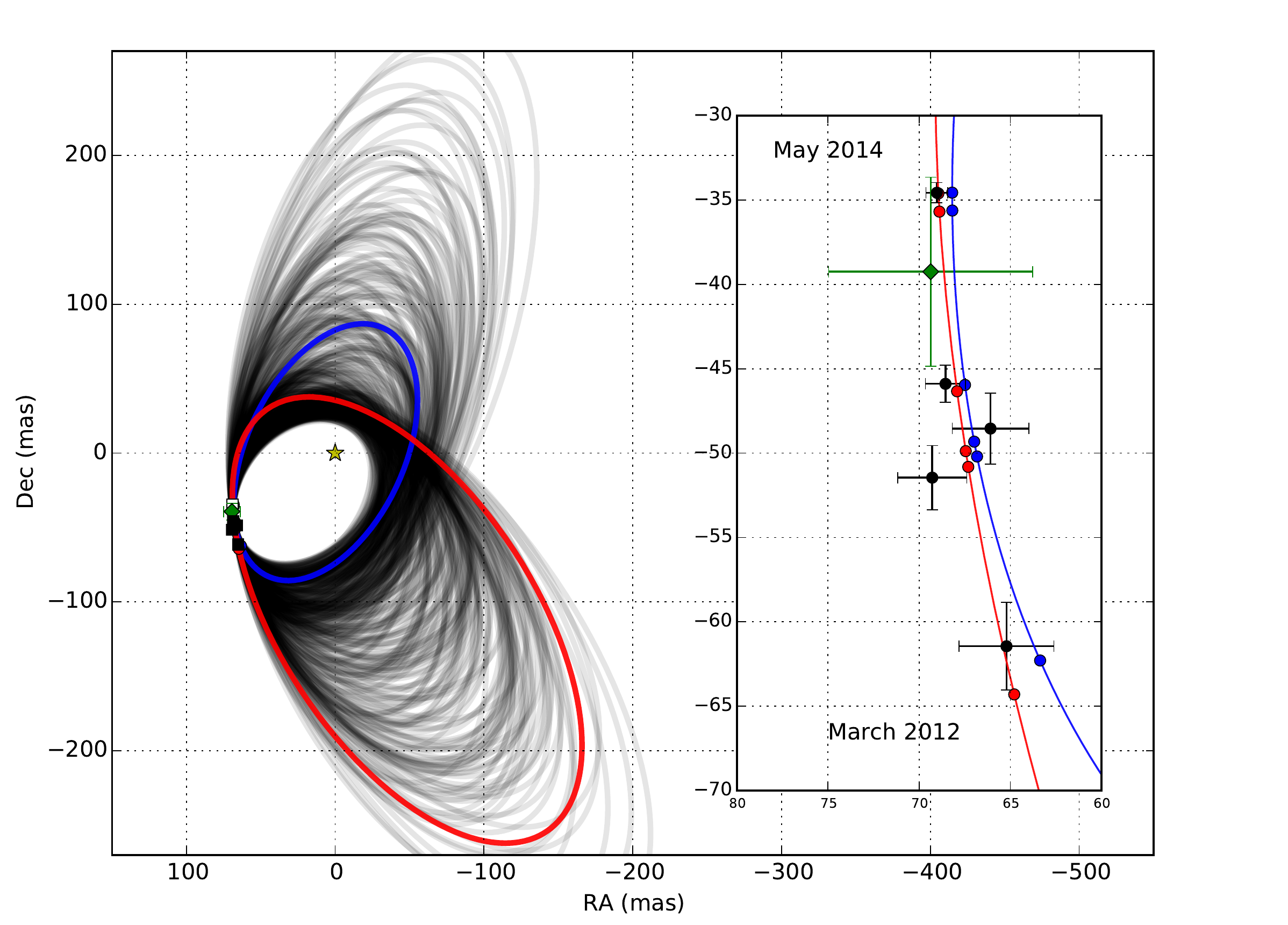}
      \caption{Likely orbits for HD142527B from MCMC simulation. The red and blue curves correspond to maximum likelihood parameters. The gray curves indicate possible solutions from the MCMC computations. The SAM observation indicate black squares. The green lozenge corresponds to a  MagAO observation reported by \citet{2014ApJ...791L..37R}.}
         \label{figTraj}
   \end{figure}

   \begin{table}
      \caption[]{HD\,142527\,B orbital observations}
         \label{tbPos}
         \centering
\begin{tabular}{l c c c}
\hline\hline
Instrument & Date & Sep (mas) & PA (deg) \\
\hline 
NACO (SAM)    &11 March 2012&$89.7\pm2.6$&$133.1\pm1.9$ \\
NACO (SAM)    &17 March 2013&$82.0\pm2.1$& $126.3\pm1.6$\\
MagAO$^{(1)}$           &11 April 2013&$86.3\pm1.9$ &$126.6\pm1.4$\\
NACO (SAM)    &14 July 2013 &$82.5\pm1.1$& $123.8\pm1.2$\\
MagAO$^{(2)}$            &April 2014&$79.7\pm5.6$&$119.5\pm8.7$ \\
GPI  (Direct)$^{(2)}$       &25 April 2014&$88.2\pm10.1$ &$123.0\pm9.2$ \\
GPI     (NRM)    &12 May 2014&$77.2\pm0.6$&$116.6\pm0.5$ \\
\hline
\end{tabular}
\tablebib{
(1) \citet{2014ApJ...781L..30C}; (2) \citet{2014ApJ...791L..37R}
}
\end{table}

   \begin{figure*}
   \centering
  \includegraphics[height=13cm]{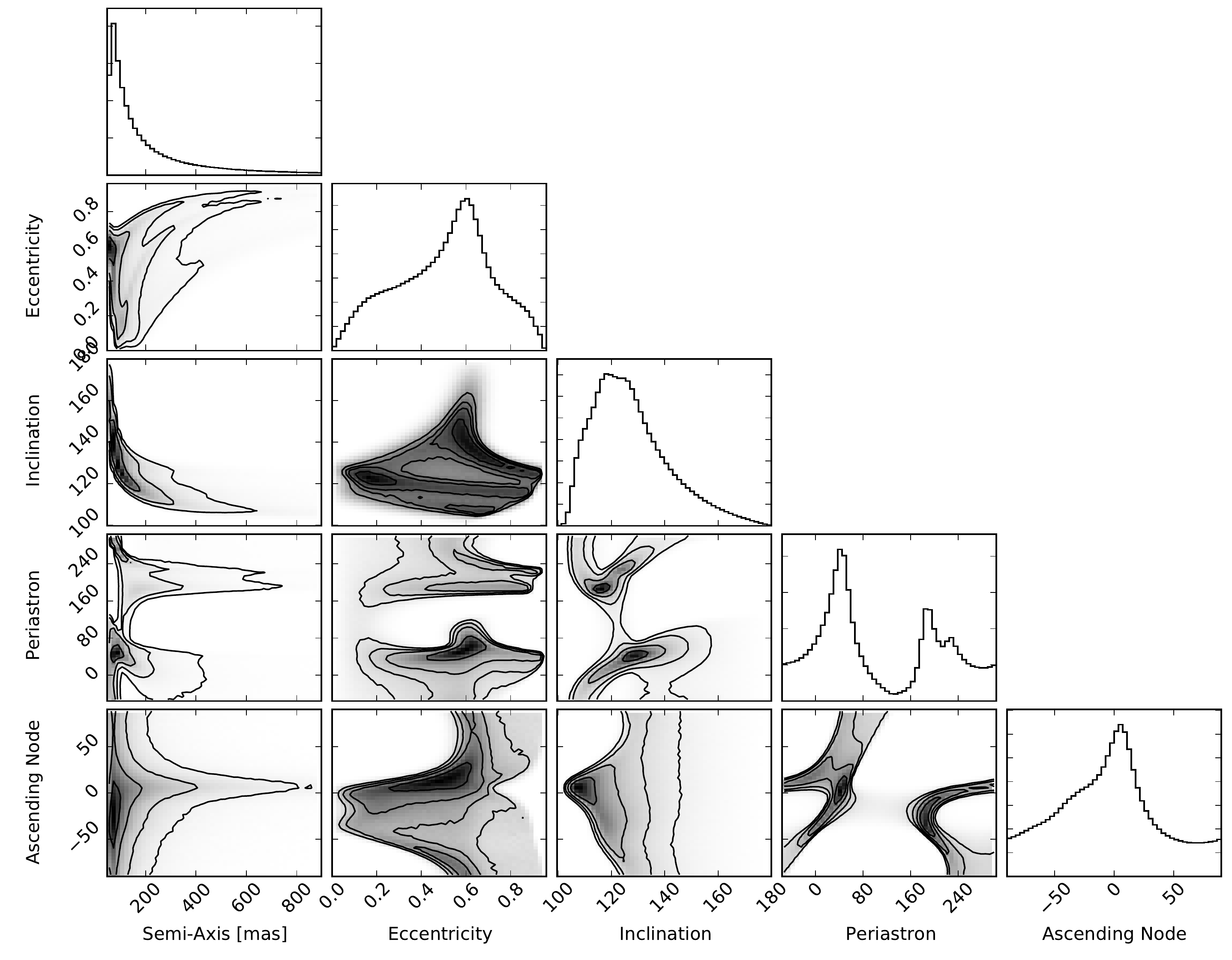}
      \caption{MCMC simulation for the parameters of the orbit of HD142527B. We assumed ${\mu} = 3.2\times 10^{-5} \,{\rm arcsec}^3 / {\rm year}^{2} $. All angles are in degrees. Ascending node is calculated for a range of values between -90 and 90 degrees. For comparison, the inner disk as simulated by \citet{2015arXiv150507732C} as a position angle (PA) for the ascending node of -3 degrees, and an inclination with respect to the plane of the sky of 43 degrees (equivalent to 137 degrees following the convention that a clockwise system has an inclination between 90 and 180 degrees).}
         \label{figMCMC}
   \end{figure*}

Known separations and position angles of HD142527B with respect to
HD142527A are presented in Table~\ref{tbPos}. This includes all SAM
observations, plus two MagAO observations
\citep{2014ApJ...781L..30C,2014ApJ...791L..37R} and one GPI
polarimetric observation \citep{2014ApJ...791L..37R}. We included the two MagAO observations in
our analysis, but did not include the GPI
polarimetric observation (obtained through direct imaging).  This GPI observation is far off the other measurements, possibly because of scattered emissions that biased the determination of the barycenter.
   
This dataset covers two years of observations, but only a small arc of
the orbit is observed, corresponding to 15 degrees in position angle.
The first approximation is therefore to consider that our
knowledge is limited to a position and a velocity vector,
projected on the plane of the sky. Thus, the position and velocity 
orthogonal to the plane of the
sky remain unknown and prevent us from fully characterizing this orbit.
However, we can still place constraints on some of the orbital parameters.
From our data listed in Table~\ref{tbPos}, we computed a vector position
$(x,y)=(68.0\pm0.7,-47)$\,mas and a vector speed $( \dot x, \dot
y)=(2.0\pm0.9,13.4\pm0.9)$\,mas/yr. The position in right ascension is
labeled $x$ and $y$ for the declination. The $y$ value was
arbitrarily fixed to $-47\,$mas to force the determination of the true
anomaly at this declination (the choice of this fixed parameter does not affect the orbital elements). Finally, we defined ${\mu}$ as a scaled mass
parameter that characterizes the acceleration at any given angular
separation,
\begin{equation}
{\mu} = GM/d^3 = 4\pi^2\left(\frac{M}{M_{\sun} } \right) \left(\frac{d}{pc}\right)^{-3} \ {\rm arcsec}^3 / {\rm year}^2 \,,
\end{equation}
where $G$ is the gravitational constant. We used $M=2.2\,M_{\sun}$ as the total mass of the binary system and $d=140$\,pc for the distance of the binary system. Hence, for the HD142527AB system
\begin{equation}
{\mu} = 3.2\times 10^{-5} \,{\rm arcsec}^3 / {\rm year}^{2} \,.
\end{equation}

The assumption that the companion is gravitationally bound to the main
stellar object (eccentricity below or equal to 1) allows us to delimit
a parameter space for the $z$ and $\dot z$ components. This $(z, \dot
z)$ parameter space is illustrated in Fig.~\ref{z_zdot} with dashed
lines. The contours represent the six different parameters of the
orbit. They correspond to the semimajor axis $a$ in mas, the
eccentricity $e$, the inclination $i$ relative to the sky plane, the
position angle of the ascending node, the argument of the periapsis,
and the true anomaly at which the companion reaches $y = -47\,$mas. All
elements are derived following the methodology described by
\citet{2015MNRAS.448.3679P}. 

Without any prior knowledge or probabilistic knowledge on $z$ and $\dot z$,  Fig.~\ref{z_zdot} can be interpreted that any semimajor axis, eccentricity, or inclination within the dashed contours is possible. For example, the semimajor axis can go from 52\,mas to infinity.

\subsection{Full determination of the orbital parameters}



We also investigate possible orbits using an MCMC analysis as an alternative to the small arc technique presented in the previous section. Such an alternative technique has already been used for small arc dataset \citep[e.g., on Fomalhaut by][]{2013ApJ...775...56K}. This method also has the advantage to better benefit from individual observational errors over multiple observations.

We used the MCMC python library emcee \citep{2013PASP..125..306F}. We computed the walkers for the variables
   $(x, z,\dot x, \dot y, \dot z)$ and converted them to a probability distribution
   function for the orbital parameters. Any corresponding orbital element with an eccentricity
   above 1 was given a probability of zero.  Within this boundary condition, the $z$ and $\dot z$ are given uniform prior distributions. The position angle of
   the ascending node was forced to lie between -90 and 90
   degrees. The relation between $(x, y, z,\dot x, \dot y, \dot z)$
   and the orbital parameters are given in Appendix B of
   \citet{2015MNRAS.448.3679P}.

   The results from the MCMC simulation are presented in
   Fig.~\ref{figMCMC}. Unfortunately, the small range of position angles
   covered by our dataset does not strongly constrain the orbital elements. Within 1 sigma probability (68\% of the results),
   they are the following: a semimajor axis $a = 140^{+120}_{-70}
   \,$mas, an eccentricity $e =0.5 \pm 0.2$, an inclination of $i =
   125 \pm 15$ degrees, and an angular position of the right ascending
   node of $-5 \pm 40$ degrees.

   The angle of the periastron is the least constrained
   parameter. Two distinct families of orbits 
   emerged from this simulation: one with a periastron
   around 40 degrees and the other at a periastron around 200
   degrees. These two families of orbits are presented
   in Fig.~\ref{figTraj}. In this figure, all the gray curves have
   equal probabilities. The blue and red curves correspond to a
   maximum of probability. The first family, shown with the blue
   curve, corresponds to a trajectory where the companion has just
   passed the periastron. The second family, shown with the red
   curve, corresponds to a trajectory where the companion is going to
   pass the periastron.

\section{Discussion}
\label{secCon}

\subsection{Summary}

\citet{2012ApJ...753L..38B} observed a NIR excess around
HD142527B.
Without a proper estimation of the level of NIR emission from the circumsecondary disk, it was difficult to be conclusive about the mass.
In this work, we extended our observational baseline to the $M$ band, but more importantly to the $J$ band. We were able to model the NIR emission and conclude that the $J$ band was mostly unaffected by the circumsecondary emission (to the 10\% level).

If we assume that the NIR emission is caused by a circumsecondary environment, we can
focus on the shorter wavelength observations ($R$ and $J$ band observations) to determine the parameters of the star. 
The result is that HD142527B is
 a fairly standard young low-mass star that matches standard evolution mechanisms. However, at least according to  the \citet{2015A&A...577A..42B} models, this last
statement is only valid if we assume an age of $1.0^{+1.0}_{-0.75}\,$Myr.

\subsection{Age of the system}


If we increase the age of the system to $5$\,Myr \citep[according to][]{2014ApJ...790...21M}, then the colors of
the companion start to disagree with its absolute magnitudes. To reconcile a more
advanced age with our observations, we would have to
increase the apparent magnitude (make the star fainter). 
For the magnitudes to agree
in the HR diagrams of Fig.~\ref{figEv} with an age of $5 \pm
1.5$\,Myr, we would have to add 1.0 and 1.5\,mag to the $R$ and
$J$ band magnitudes, respectively.  That would put the target at 70\,pc, which is not
compatible with the distance estimated in Section~\ref{secdist}. Dust absorption would only decrease the absolute magnitude,
giving an opposite effect.


Although previous episodes of intense accretion can change the structure and thus the position 
of a young object in a magnitude-temperature diagram, it seems difficult to invoke accretion effects 
to get an age of 5 Myr, instead of 1 Myr, for the observed luminosity of the low-mass companion.
As discussed in \citet{2009ApJ...702L..27B,2012ApJ...756..118B},
cold accretion would have the opposite effect. The only, very unlikely possibility, would 
be for the low-mass star to have had a previous episode of intense hot accretion that would increase its luminosity and radius. 
The object would thus look younger than its nonaccreting counterpart.
This accretion episode, however, should be recent. Otherwise, the object would have time to contract back to a size compatible with the models.

The age estimations of HD142527 A\&B rely on  evolutionary models that have their own intrinsic source of errors. It has been shown that these uncertainties can be well above a few million years \citep{2014prpl.conf..219S}. Especially, it has been shown that the ages of intermediate-mass stars tend to disagree with the ages of the T Tauri stars in same
clusters \citep{2003ApJ...585..398H}. 

This system thus seems to confirm the existence of a disagreement between ages derived from low-mass star models and those from
intermediate-mass star models.  Since the source of uncertainties in the physics and modeling of these two families of objects 
is different, more efforts and more of those systems are needed to determine whether the remaining uncertainties are inherent to the former or the latter models (or to both)

\subsection{Interaction with the circumprimary and circumbinary disks}

   \begin{figure}
   \centering
  \includegraphics[height=7cm]{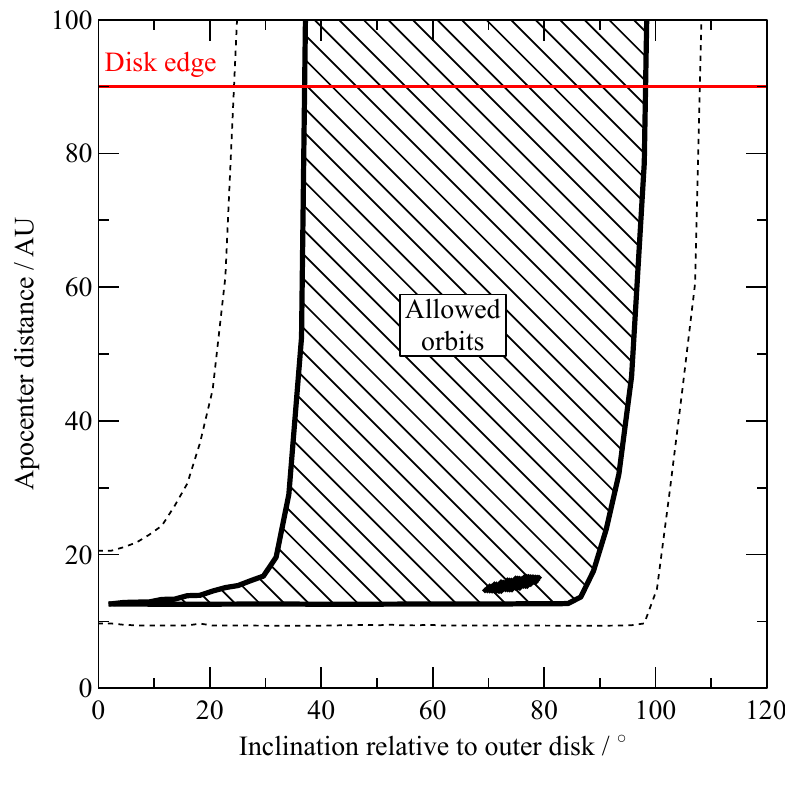}
  \caption{ Possible apocenter distance of the companion vs. its
    inclination relative to the outer disk plane. The shaded region
    shows allowed orbits and the dashed lines indicate the $1\sigma$ errors on
    this region. The thick red line is the inner edge of the outer
    disk, showing that the companion cannot have an apocenter close to
    this edge and simultaneously orbit in the same plane as the
    disk. The black area in the lower right of the allowed region
    shows orbits lying within 5 degrees of the inner disk plane.  }
         \label{figTim}
   \end{figure}

   The orbital elements of the binary system are not in agreement with the orientation and inclination of the outer disk, but they are
in agreement with the
   parameters of the inner disk derived by \citet{2015ApJ...798L..44M}
   and \citet{2015arXiv150507732C}: an inclination of 140 degrees with
   respect to the sky plane (70 degrees with respect to the outer
   disk) and a PA of the ascending node at -3\, degrees. It is
   therefore likely that the kinematics of the inner disk are linked to
   the kinematics of HD142527B.

The remaining question concerns the gap. Is the inner edge of the
 outer disk truncated by the companion? If there are orbital
   solutions that place the companion in the outer disk plane with
   its apocenter located close to the inner edge of the disk, then the
   companion may have an ongoing role in sculpting the
   disk. Figure~\ref{figTim} shows the apocenter distance
of the companion   versus its inclination to the outer disk plane for all possible
   orbital solutions. We assumed the outer disk is circular with
   an inner edge at 90 AU, an inclination of 28 degrees to the sky, and
   a major axis position angle of 160 degrees
   \citep{2011A&A...528A..91V,2015ApJ...798...85P}. The figure shows
   that for the apocenter distance of the companion to be comparable to the
   inner edge of the outer disk, its inclination to the outer disk
   plane must be greater than $\sim 30^\circ$. Furthermore if we
   assume that the companion lies in the plane of the inner disk (as
   suggested above) then its apocenter is only around 15 AU, and it is
   misaligned to the outer disk by $\sim 70^\circ$. Hence the
   companion cannot simultaneously lie in the outer disk plane and
   have an apocenter comparable to the inner edge of the outer disk, so
   it is unlikely that the companion is responsible for truncating the
   outer disk.  


\begin{acknowledgements} SL acknowledges fruitful discussions with S. Casassus about the existence of HD142527B and the inner disk of HD142527A. This research made use of Astropy, a community-developed core Python package for Astronomy \citep{2013A&A...558A..33A}.
This work was supported by the French National Agency for Research (ANR-13-JS05-0005) and the European Research Council  (ERC-STG-639248). AG and AS acknowledge support from NSF Graduate Research Fellowship grant no. DGE-1232825 and NASA grant NNX11AF74G. JO acknowledges support from the Millennium Nucleus RC130007 (Chilean Ministry of Economy). IB acknowledges the European Research Council through grant ERC-AdG No. 320478-TOFU.
Based on observations collected   at   the   European   Southern
Observatory (ESO) during runs 
088.C-0691(A),
090.C-0649(A),
        091.C-0572(A), and
        094.C-0608(A).  
Also based on observations obtained at the Gemini Observatory (programs GS-2014A-SV-406 and GS-ENG-GPI-COM), which is operated by the 
Association of Universities for Research in Astronomy, Inc., under a cooperative agreement 
with the NSF on behalf of the Gemini partnership: the National Science Foundation 
(United States), the National Research Council (Canada), CONICYT (Chile), the Australian 
Research Council (Australia), Minist\'{e}rio da Ci\^{e}ncia, Tecnologia e Inova\c{c}\~{a}o 
(Brazil) and Ministerio de Ciencia, Tecnolog\'{i}a e Innovaci\'{o}n Productiva (Argentina).
 \end{acknowledgements}

\bibliographystyle{aa}   
\bibliography{HD142} 

\end{document}